\documentclass[12pt,a4paper,final]{iopart}

\usepackage{iopams}  
\usepackage{graphicx}
\expandafter\let\csname equation*\endcsname\relax
\expandafter\let\csname endequation*\endcsname\relax
\usepackage{amsmath}
\usepackage{bm}
\usepackage{anyfontsize}
\usepackage{xcolor}

\definecolor{forestgreen(traditional)}{rgb}{0.0, 0.27, 0.13}
\definecolor{mordantred19}{rgb}{0.68, 0.05, 0.0}
\definecolor{ultramarine}{rgb}{0.07, 0.04, 0.56}
\definecolor{purple(html/css)}{rgb}{0.5, 0.0, 0.5}

\usepackage[multiple,bottom]{footmisc}
\makeatletter
\renewcommand\@makefntext[1]%
     {\noindent\makebox[0pt][r]{\textsuperscript{\@thefnmark}\,}#1}
\makeatother
\makeatletter
\renewcommand\@makefnmark[1]%
     {\noindent\makebox[8pt][r]{\textsuperscript{\@thefnmark}\,}#1}
\makeatother
\makeatletter
\let\@fnsymbol\@arabic
\makeatother
\usepackage[hyperfootnotes=true]{hyperref}
\hypersetup{colorlinks=true, linkcolor=ultramarine, urlcolor=ultramarine, citecolor=ultramarine}

\begin{document}

\title{Evidence for the existence of a novel class of supersymmetric black holes with AdS$_5\times$S$^5$ \\ asymptotics}

\author{Julija Markevi\v{c}i\={u}t\.e and Jorge~E.~Santos}
\address{Department of Applied Mathematics and Theoretical Physics, University of Cambridge, Wilberforce Road, Cambridge CB3 0WA, UK}
\eads{\href{mailto:jm722@cam.ac.uk}{jm722@cam.ac.uk}, \href{mailto:jss55@cam.ac.uk}{jss55@cam.ac.uk}}

\begin{abstract}

\noindent We construct a new class of charged, rotating hairy black holes in a consistent truncation of $\mathcal{N} = 8$ supergravity, which retains one charged scalar field and a U$(1)$ gauge field. These hairy solutions can be uplifted to solutions of type IIB supergravity with AdS$_5\times$S$^5$ asymptotics.  We find rotating hairy black holes with finite entropy arbitrarily close to the supersymmetric bound - the resulting supersymmetric solution is a one-parameter extension of the Gutowski-Reall solution. These solutions have finite curvature invariants (including at extremality), but in the extremal limit exhibit diverging tidal forces in the near horizon region. Nevertheless, we argue that these limiting supersymmetric black holes can be consistently studied within the supergravity approximation.

\vspace{+1em}
\end{abstract}



\section{Introduction}
Gauge/gravity duality \cite{Maldacena:1997re,Gubser:1998bc,Witten:1998qj,Aharony:1999ti} is one of the cornerstones of modern high-energy physics. It has far reaching implications for a wide range of fields, including higher-dimensional gravity, numerical relativity, lattice simulations, string theory, quantum field theory, confinement and condensed matter.

The reason why AdS/CFT is useful is also the reason why it is difficult to prove. It is a strong-weak duality, meaning that when one of the sides of the correspondence is (in principle) solvable, the other is in a regime where no generic known technique can be used to study it. In its original form, it conjectures an equivalence between four-dimensional $\mathcal{N}=4$ Super-Yang-Mills (SYM) with gauge group SU$(N)$ and coupling constant $g_{\mathrm{YM}}$ and ten-dimensional type IIB string theory with AdS$_5\times$S$^5$ asymptotics, string coupling $g_s$  and string length $\ell_s \equiv \sqrt{\alpha^\prime}$~\cite{Maldacena:1997re}.  Furthermore, the field theory lives on the conformal boundary of AdS$_5$, which in global coordinates is the Einstein static Universe $\mathbb{R}_t\times$S$^3$. The free parameters on each side of the correspondence are identified via
\begin{equation}
g^2_{\mathrm{YM}}=2\pi g_s\qquad\mathrm{and}\qquad 2\,\lambda\,=L^4/{\alpha^\prime}^2\,,
\label{eq:string}
\end{equation}
where $L$ is the radius of curvature of AdS$_5$ and the t'Hooft coupling is defined as $\lambda\equiv g^{2}_{\mathrm{YM}}N$. The string theory side is mostly understood in the limit $g_s\ll1$ and in the supergravity limit $L^2/\alpha^\prime\gg1$. This can be achieved by simultaneously taking  $N\to+\infty$ and $\lambda \gg1$. Even in this limit, when the string theory side reduces to a classical supergravity calculation, the duality is hard to study, since the field theory side is strongly coupled.

There are, however, many nontrivial tests of the correspondence in this limit, some of which have been reviewed in \cite{Aharony:1999ti}. One of the major successes of string theory in flat space~\cite{Strominger:1996sh} is a microscopic counting of the black hole entropy from first principles. However, this has not yet been accomplished in the original form of AdS/CFT just stated above\footnote{Note however that remarkable progress has been made in counting the entropy of black holes dual to certain phases of topologically twisted deformation of ABJM theory \cite{Benini:2015eyy,Benini:2016rke}.}. On the gravity side, Gutowski and Reall constructed a supersymmetric black hole in \cite{Gutowski:2004ez}, whose entropy has never been accounted for in $\mathcal{N}=4$ SYM. These black holes are 1/16 BPS solutions, and in its original form, have two equal magnitude angular momenta in AdS$_5$ and carry three equal magnitude R-charges. They were originally found within five-dimensional, minimal, $\mathcal{N}=1$ gauged supergravity, and were readily uplifted to type IIB string theory with AdS$_5\times$S$^5$ asymptotics using the results contained in \cite{Gauntlett:2004cm}. A surprising result of \cite{Gutowski:2004ez} is the fact that the black hole that was found has a single free parameter, \emph{i.e.} corresponds to a one parameter family of solutions. This is surprising because the most general 1/16 BPS state in $\mathcal{N}=4$ SYM (after imposing equality of the R-charges and equal magnitude angular momentum) can be shown to depend on two real fugacities. This is turn implies, that if we were to attempt a microscopic accounting of the entropy using the field theory we would have to stop almost immediately, because the number of free parameters characterising $1/16$ BPS states (within our class of symmetries) do not match on both sides of the correspondence \cite{Kinney:2005ej}. In this letter, we attempt to shed light on this problem, by giving evidence in favour of a new family of supersymmetric black hole solutions that depend on two free parameters, thus generalising the original Gutowski-Reall black hole.

In this Letter we proceed as follows. In section \ref{sec:action} we present the action that we used throughout our work. Next, in section \ref{sec:results}, we demonstrate the existence of a new family of hairy black hole solutions which approach the BPS bound. We find that such solutions have diverging tidal forces, and argue that they can be accurately described within the supergravity approximation. We conclude with discussion and future directions in section \ref{sec:conclusions}.

\section{\label{sec:action}The action}

Due to its complexity, we will not work directly in type IIB supergravity, but instead we will focus on five-dimensional $\mathcal{N}=8$ supergravity, since the latter is a consistent truncation of the former \cite{deWit:2013ija,Godazgar:2013pfa,Godazgar:2013nma,Godazgar:2013dma,Godazgar:2013oba,Lee:2014mla,Baguet:2015sma} and appears much more tractable. In fact, we will be working with a further reduction of $\mathcal{N}=8$, which was first presented in \cite{Bhattacharyya:2010yg} and whose static black hole solutions were studied in a great detail in \cite{Markeviciute:2016ivy}. These solutions can be seen as a string theory embedding of the global black holes found in \cite{Dias:2011tj}. From the five-dimensional perspective, this theory contains a metric $g$, a U$(1)$ gauge field $A$ and a complex charged scalar field $\Phi$ that minimally couples to $A$. The scalar field, however, will have a very non-minimal coupling to gravity. 

Our action reads
\begin{align}
\begin{split}
\label{eq:action}
 S=\frac{1}{16\pi G_5}\int \mathrm{d}^5x\sqrt{-g}\left\{R+12-\frac{3}{4}F_{ab}F^{ab}-\frac{3}{8}\left[(D_a\Phi)(D^a\Phi)^*-\frac{\nabla_a\lambda\,\nabla^a\lambda}{4(4+\lambda)}-4\lambda\right]\right\}\\
-\frac{1}{16\pi G_5}\int F\wedge F\wedge A,
 \end{split}
\end{align}
where $\lambda = \Phi \Phi^*$, ${}^*$ denotes complex conjugation, $F=\mathrm{d}A$, $D=\nabla-i\,e A$. We have set the AdS$_5$ length scale $L$ to unity and $G_5=\pi/(2N^2)$. The complex scalar field $\Phi$ has electric charge $e=2$ and mass square $m_\Phi^2=-4$, saturating the five-dimensional Breitenl\"ohner-Freedman (BF) bound~\cite{Breitenlohner:1982jf}. When $\Phi$ vanishes, this theory reduces to minimal gauged supergravity, where the black holes of \cite{Gutowski:2004ez} were initially found. The equations of motion derived from this action can be found in the \ref{sec:appendixA}. The supersymmetric black hole solutions of \cite{Gutowski:2004ez} later emerged as a particular extremal limit of finite-temperature solutions \cite{Cvetic:2004ny,Chong:2005hr}. One expects the latter to be unstable to charged superradiance \cite{PhysRevLett.101.031601,Dias:2010ma,Dias:2011tj,Markeviciute:2016ivy} when considering cold black holes, thus prompting the existence of hairy solutions where $\Phi\neq0$. These are the solutions whose supersymmetric limit we aim to discuss in the manuscript.

The most general \emph{ansatz} for stationary, equal magnitude angular momenta, asymptotically AdS$_5$ black hole solutions with spherical horizon topology in an arbitrary gauge is
\begin{equation}
 \label{eq:ansatzr}
 \mathrm{d}s^2=-f(r)\mathrm{d}t^2+g(r)\mathrm{d}r^2+\\
 \Sigma(r)^2 \left[h(r)\left(\mathrm{d}\psi+\frac{1}{2}\cos{\theta}\mathrm{d}\phi-w(r)\,\mathrm{d}t\right)^2+\frac{1}{4}\mathrm{d}\Omega^2_2\right]
\end{equation}
which has co-homogeneity one. When $f(r)=g(r)^{-1}=1+r^2$, $\Sigma(r) = r$, $w(r)$ and $h(r)=1$ we recognise the line element above as that of AdS$_5$ in global coordinates, where the round 3-sphere is written as a Hopf fibration. The fiber is parametrised by the coordinate $\psi$ with a period $2\pi$, and $\theta, \phi$ are the standard polar coordinates on~S$^2$. The level surfaces of~(\ref{eq:ansatzr}) are homogeneously squashed S$^3$. The known solutions to the equations of motion derived from (\ref{eq:action}) have been found in the radial gauge where $\Sigma(r) = r$, and we will also work in this gauge. The conformal boundary is thus located at $r=+\infty$, and black hole horizon is the null hypersurface $r=r_+$, where $f(r)$ vanishes linearly and $g(r)$ has a simple pole.

 The gauge field \emph{ansatz} compatible with the symmetries of the metric is
 \begin{equation}
 A=A_t(r)\,\mathrm{d}t+A_\psi (r)\,\left(\mathrm{d}\psi+\frac{1}{2}\cos{\theta}\,\mathrm{d}\phi\right),
 \label{eq:gauge}
 \end{equation}
and we also set ${\Phi=\Phi^*=\Phi(r)}$, which partially fixes the residual U$(1)$ gauge symmetry.

This \emph{ansatz} has a residual diffeomorphism gauge freedom associated with shifts along $\psi$
\begin{equation}
\psi\rightarrow\psi +\alpha\,t,\qquad w\rightarrow w+\alpha,
\label{eq:gaugeres}
\end{equation}
for some constant $\alpha$. This can be used to set $w\rightarrow 0$ at the conformal boundary, enforcing a frame with the conformal boundary $\mathbb{R}_t\times$S$^3$. $w(r)$ evaluated at the horizon is then identified as the black hole angular velocity $\Omega$.

At the conformal boundary we want our solution to approach $\mathbb{R}_t\times$S$^3$ at the appropriate rates \cite{Ashtekar:1984zz,Henneaux:1985tv,Henningson:1998gx,deHaro:2000vlm}
\begin{align}
\begin{split}
f(r)&=r^2+1+\frac{C_f}{r^2}+\mathcal{O}(r^{-4}),\quad  A_\psi(r)=\mathcal{O}(r^{-2}),\quad A_t(r)=\mu+\frac{2\,q}{r^2}+\mathcal{O}(r^{-6}) \,,
\\
g(r)&=\frac{1}{r^2}-\frac{1}{r^4}+\mathcal{O}(r^{-6}),\quad w(r)=\frac{2\,j}{r^4}+\mathcal{O}(r^{-5})\,,
\\
h(r)&=1+\frac{C_h}{r^4}+\mathcal{O}(r^{-6})\,,\quad\Phi(r)=\frac{\varepsilon}{r^2}+\mathcal{O}(r^{-4})\,,
\label{eq:expansion}
\end{split}
\end{align}
where we used standard quantisation for the scalar field $\Phi$. As detailed in the~\ref{sec:appendixA}, $J\equiv j\,N^2$ is the angular momentum of the solution, $Q\equiv q\,N^2$ is the total charge, $\varepsilon$ is directly proportional to the expectation value of the operator dual to $\Phi$, and $M\equiv m\,N^2=(C_f-3\,C_h)\,N^2/4$ measures the total energy. In this theory, supersymmetric solutions satisfy
\begin{equation}
M = 2\,J + 3\,Q\,.
\label{eq:susy}
\end{equation}
We will give evidence of a new 2-parameter family of solutions that satisfy such bound and where $\Phi$ is non-vanishing and finite. Using the first law of black hole mechanics
\begin{equation}
\mathrm{d}M=T\,\mathrm{d}S+2\Omega\,\mathrm{d}J+3\,\mu\,\mathrm{d}Q
\end{equation}
and Eq.~(\ref{eq:susy}) one concludes that $\Omega=\mu=1$\footnote{Perhaps amusingly, the fact that these black holes have $\Omega=1$ means they are nonlinearly unstable to \emph{rotating} superradiance, since it has been shown in \cite{Green:2015kur} that any black hole with $\Omega>1$ is superradiantly unstable to perturbations that break the rotating symmetry generated by $\partial_\psi$.} and $T=0$ on such limiting configurations (assuming $S\neq0$ in the limiting case, as we shall confirm below).

Our most general non-extremal solution will depend on the three asymptotic charges: $m$, $j$ and $q$. Alternatively, we can use horizon quantities, and formulate the problem in terms of finding new solutions for given values of the Hawking temperature $T$, black hole angular velocity $\Omega$ and chemical potential $\mu$. For numerical convenience we will generate solutions by dialing different values of the scalar field $\Phi$ at the horizon, $\varepsilon_H$, while holding $j$ fixed and decreasing the black hole horizon size $r_+$. 

In order to solve the resulting equations of motion we use a standard Newton-Raphson relaxation routine on a single Chebyshev grid\footnote{We can show that there are no non-analytic terms at either end of our integration domain, and as such we have exponential convergence as we approach the continuum.} and provide detailed convergence tests in the \ref{sec:appendixC}. For more details on the numerical implementation see for instance \cite{Dias:2015nua}.

\section{\label{sec:results}Results}

Hairy charged black holes with $j=0$ interpolate the solution space of the charged black holes between the near extremal RNAdS and the zero angular momenta BPS bound $M=3Q$ \cite{Markeviciute:2016ivy}. Furthermore, on the BPS bound they reduce to hairy, horizonless supersymmetric solutions in the $T\rightarrow 0$ and $T\rightarrow\infty$ limits. In the zero temperature limit the corresponding limiting solution is smooth, while for $T\to+\infty$ it becomes singular.

We find that the inclusion of $j>0$ leads to a very similar picture with one important difference. The hairy rotating black holes branch off the near-extremal rotating and charged black holes, and exist arbitrarily close to the BPS bound for all charges. However, the cold black hole phase retains non-zero entropy as $T\rightarrow 0$ and $j>0$. The hairy black hole solutions that we constructed numerically have $T>0$, and in this paper we are interested in analysing their near-extremal quantities such as entropy and curvature invariants. The hairy solution moduli space displays many fascinating properties, and the phase diagram is rather intricate. Extensive analysis of the full three-dimensional charged, rotating and hairy black hole solution space and its thermodynamic properties is presented in the companion manuscript \cite{Markeviciute:2018cqs}. In particular, the hairy black holes always dominate the microcanonical ensemble.
	
\begin{figure}[t]
\centering
\includegraphics[width=0.5\linewidth]{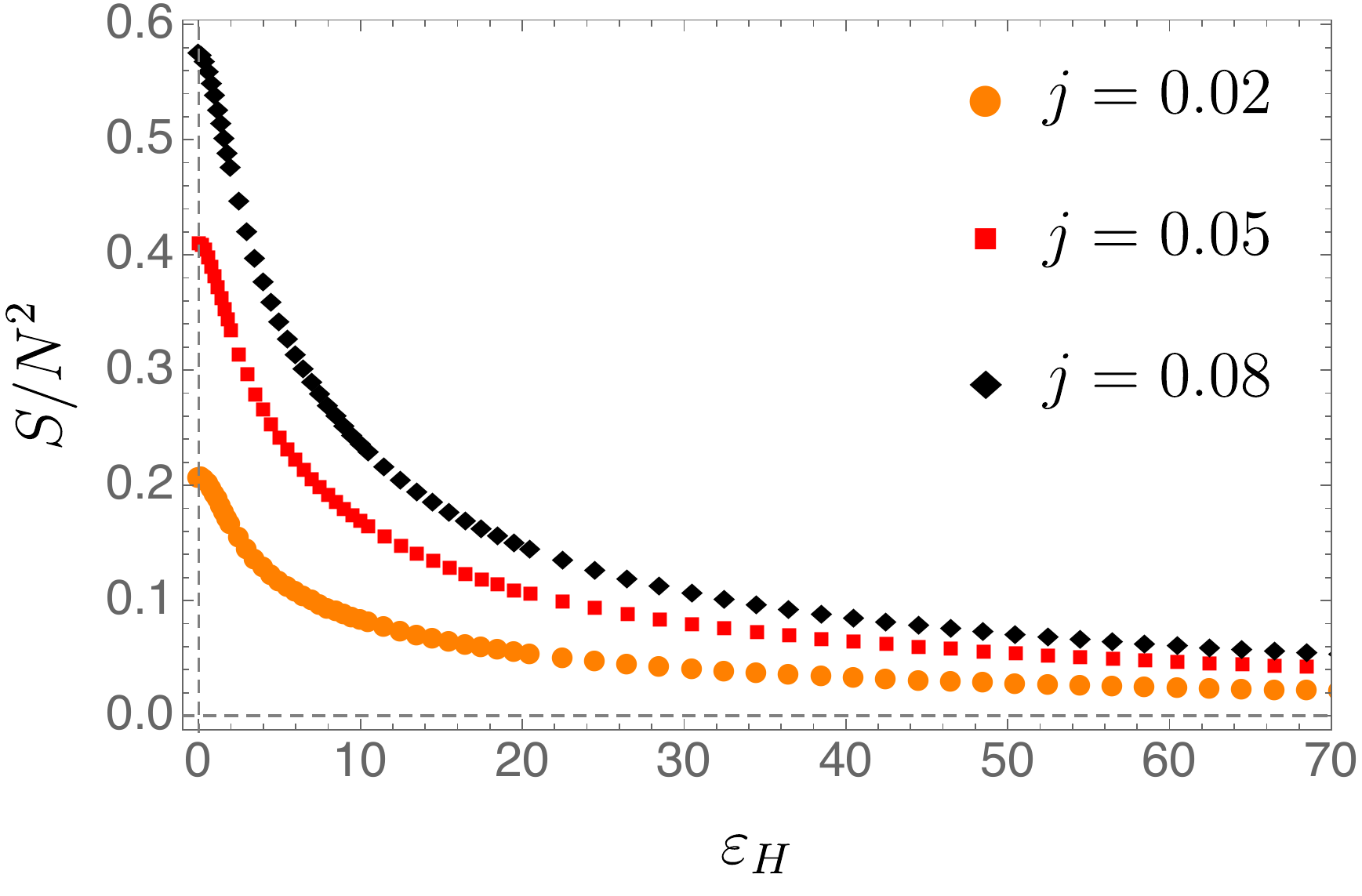}
\caption{Extrapolated zero temperature entropy of the hairy black holes with constant $j = 0.02$ (orange disks), $j = 0.05$ (red squares) and $j = 0.08$ (black diamonds) against the central scalar field $\varepsilon_H$. The variation in the entropy as we lower the temperature further gives error less than 0.1$\%$.}
\label{fig:entropy}
\end{figure}

We monitored the black hole entropy $S$ in the limit of fixed~$j$ and fixed horizon scalar $\varepsilon_H$, as the temperature decreased down to $T= 5 \times 10^{-3}$. Decreasing the temperature further at the expense of using a denser numerical grid gave only a very small further variation in $S$. For example, the entropy only changed at the $0.1\%$ level in the temperature interval of $10^{-2}$ to $10^{-3}$. The results of our numerical experiment can be found in Fig.~\ref{fig:entropy}, where we plot the extrapolated zero temperature entropy at several fixed values of $j$, as a function of $\varepsilon_H$. We find that the entropy behaves like a power law $S\propto (\varepsilon_H)^{\alpha}$ at large $\varepsilon_H$ with an exponent $1/2 < \alpha < 1$. The limiting solution has $\Omega\to1$ and $\mu\to1$ as well as satisfying Eq.~(\ref{eq:susy}) to better than 0.2$\%$ accuracy.

The existence of these black holes has been conjectured in \cite{Bhattacharyya:2010yg} based on a weakly interacting model. We have done a detailed comparison of our numerical results with those in \cite{Bhattacharyya:2010yg}, and find a good agreement when the black hole asymptotic charges are sufficiently small (where the approximation of \cite{Bhattacharyya:2010yg} is valid). For the hairy black holes with the same mass, charge and angular momentum, the agreement is better than to $0.5-5\%$. For instance, a numerical solution which has $j=0.05$, $T=0.0015$ and entropy $S =0.405$, agrees with \cite{Bhattacharyya:2010yg} at the $3.2\%$ level in entropy.

We have studied a variety of curvature invariants as we approach this limiting supersymmetric solution and we found that they are all finite. These include $R^{abcd}R_{abcd}$, $C^{abcd}C_{abcd}$, $F^{ab}F_{ab}$ and $|\Phi|^2$. We went further, and constructed a coordinate frame in which all components of the Riemann tensor of the hairy black holes are finite everywhere, in particular at the horizon, and thus all curvature scalars derived from the Riemann tensor (and its derivatives) are finite. 

\begin{figure}[!t]
\centering
    \begin{minipage}[t]{0.49\textwidth}
    \includegraphics[width=\textwidth]{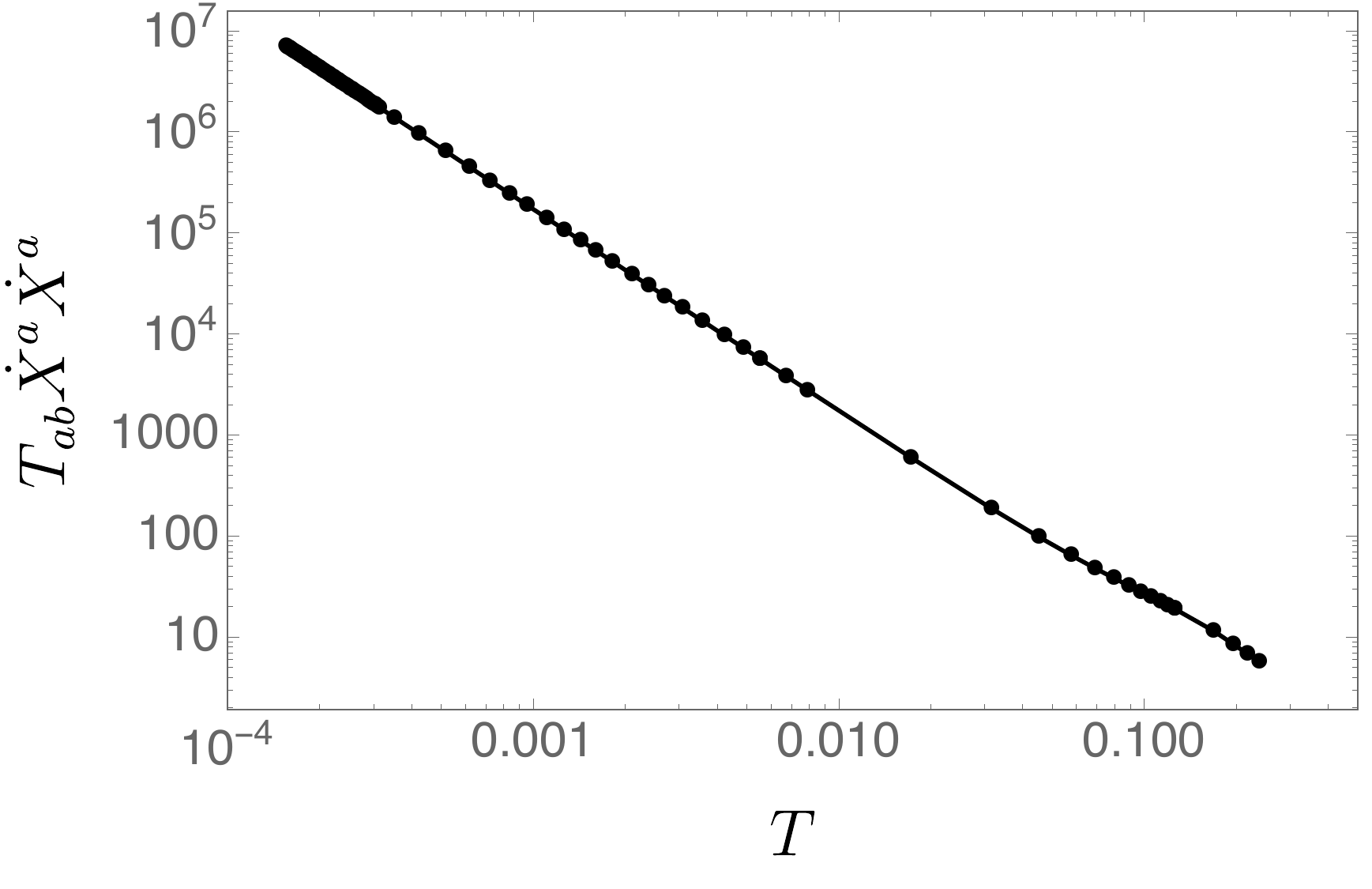}
  \end{minipage}
  \hfill
      \begin{minipage}[t]{0.49\textwidth}
    \includegraphics[width=\textwidth]{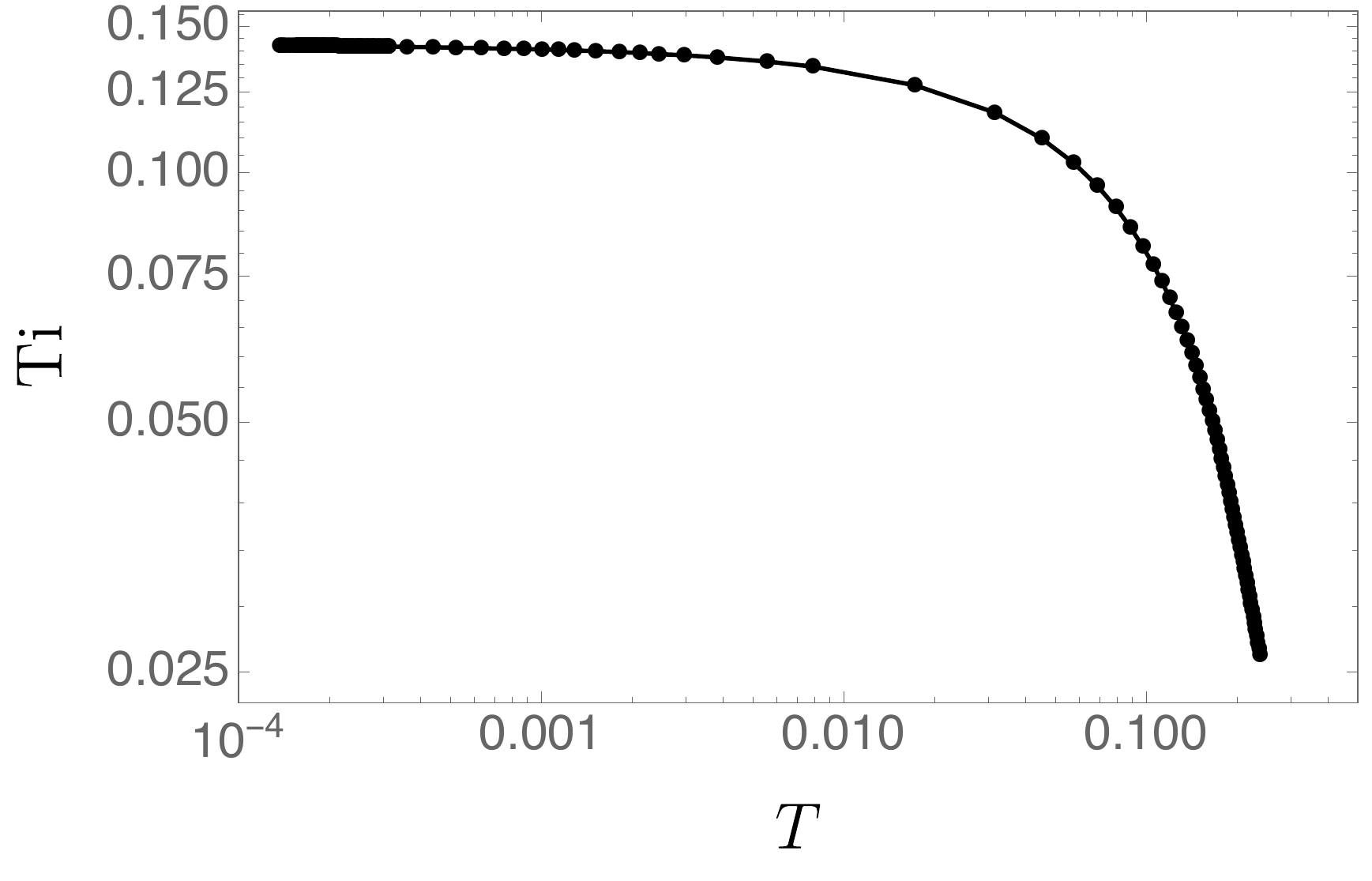}
  \end{minipage}
  \caption{\textit{Left}: The measure of tidal forces, as felt by a unit energy particle infalling along a radial geodesic, $T_{ab}\dot{X}^a\dot{X}^b$ at the hairy black hole horizon against the temperature. The quantity shown is for a black hole family with fixed $j = 0.05$ and $\varepsilon_H = 1$, and the plot is in a log-log scale. \textit{Right}: The log-log plot of the Tipler integral, for hairy black hole family with constant angular momentum $j=0.05$ and horizon scalar $\varepsilon_H=1$. The integral was computed using the $R^t_{\,\,\psi t\psi}$ component found in a PPON frame. Different components have similar qualitative behaviour.}
 \label{fig:divergence}
\end{figure}

We have also studied the tidal forces as felt by an observer infalling into the black hole. We compute the measure $T_{ab}\dot{X}^a\dot{X}^b$, where $\dot{X}^a$ is the tangent vector of a timelike ingoing geodesic parametrised by the proper time $\tau$, and $T_{ab}$ is the stress energy tensor associated with the action (\ref{eq:action}). We find that the tidal forces diverge when we approach $T=0$, as it can be observed in Fig.~\ref{fig:divergence} (\textit{left}). Furthermore, the Riemann tensor components measured in a freely falling frame diverge as we approach the extremality, confirming that there is a parallely propagated (pp) curvature singularity.

Such singularities in supergravity theories are not rare, and there have been many examples where the near extremal solutions exhibit pp singularities \cite{Gueven:2005px,Harrison:2012vy,Bhattacharya:2012zu,Brecher:2000pa,Horowitz:1997uc,Horowitz:1997ed,Horowitz:1997si,Kaloper:1996hr,Madhu:2009jh,Burko:1999zv,Narayan:2009pu}. It is worth emphasising that all our hairy black holes, including the ones just above the BPS bound are smooth solutions, and the tidal forces diverge \emph{only} at extremality. Such singularities can be regarded ``good'' in the sense of~\cite{Maldacena:2000mw,Gubser:2000nd}. We will, however, further argue the divergences exhibited by the extremal limit of our solutions can be consistently studied within the supergravity approximation as the Jacobi fields remain bounded.

The notion of strong curvature singularities was first introduced by~\cite{Ellis1977}, which were defined by the ability to crush any objects passing through the singularity to zero volume. This idea was made more precise by~\cite{TIPLER19778,CLARKE1985127}, where the strong curvature singularities were defined such that an extended object falling into the singularity retains a non-zero volume. The physical dimension of the object is defined by  linearly independent spacelike (vorticity-free) Jacobi fields along the timelike geodesic which extends to the singularity. The weak curvature singularity is defined in a similar way. In such singularities, even though the tidal forces diverge, it is still possible to have the overall effect on the volume deformation to be bounded.

The necessary and sufficient conditions for the singularity to be weak were given by Clarke and Krolak~\cite{CLARKE1985127}, and Tipler~\cite{TIPLER19778}. If the singularity resides at an affine parameter $\tau_\star$ and the particle is at rest at $\tau=0$, then for a timelike geodesic meeting the singularity at $\tau_\star$, the integral of the components of the Riemman tensor in a parallely propagated orthonormal frame (PPON)
\begin{equation}
\mathrm{Ti}=\int_{\tau_0}^{\tau_\star}\int_{\tau_0}^{\tau '}\left|R^i_{\phantom{i}0j0}(\tau'')\right|\mathrm{d}\tau''\mathrm{d}\tau'
\label{eq:Tipler}
\end{equation}
\noindent will not diverge, for any $\tau_0\in[0,\tau_\star)$. In the \ref{sec:appendixB} we present our choice of PPON adapted to our symmetries. The finiteness of $\mathrm{Ti}$ guarantees that the volume defined by the Jacobi fields remains non-zero when crossing the singularity, and thus the curvature singularity is weak.

In addition to the volume being non-zero as the object crosses the singularity, one might require the norm of the Jacobi fields themselves to remain finite~\cite{Nolan:1999tw,PhysRevD.61.064016,Nolan:2000rn}, so that there is no divergent distortion in any of the directions. The convergence of~(\ref{eq:Tipler}) implies that the Jacobi fields themselves are bounded~\cite{PhysRevD.61.064016}. We find that the Tipler integral (\ref{eq:Tipler}) remains finite, as can be seen in Fig.~\ref{fig:divergence} (\textit{right}). For different values of $j$ and $\varepsilon_H$ the curves quantitatively change, but otherwise behave in the same qualitative manner. We find that finiteness of the Jacobi fields should be more than enough to control the classical and quantum propagation of strings.

There are situations where $\mathrm{Ti}$ diverge and yet, stringy perturbation theory seems to be under good control \cite{Bao:2012yt}. It is also worth mentioning that our limiting solutions have bounded $L^2$ curvature norm, and thus can be continued as solutions of the Einstein equation past the black hole horizon \cite{Klainerman:2012wt}\footnote{Consequently, they are also weak solutions in the sense of Christodoulou \cite{Christodoulou:2008nj}.}. We believe that the limiting solutions found here behave very much like the extremal black holes of \cite{Dias:2011at}. 

Of particular interest is the near-horizon geometry of the extremal hairy black hole limit. Numerical results suggest a non-trivial, exotic geometry which may be of Lifshitz~\cite{Kachru:2008yh}, or hyperscaling violating type~\cite{Huijse:2011ef}. While the attractor mechanism in ungauged supergravity~\cite{Ferrara:1995ih,Strominger:1996kf,Ferrara:1996dd,Ferrara:1996um,Chamseddine:1996pi,Kallosh:1996vy,Ferrara:1997tw} is generally well understood, it has not been completely resolved in the five-dimensional gauged supergravity (\textit{e.g.} see \cite{Hosseini:2017mds}). We note that the entropy of the hairy BPS black holes is determined solely by the black hole charges, and therefore it is consistent with the attractor mechanism. This is a regime which is difficult to approach numerically, but should in principle be tractable analytically by utilising the BPS equations.

\section{\label{sec:conclusions}Conclusions and future directions}

Our results unveil a myriad of opportunities for future work. First, it is conceivable that our hairy black hole might not be the entropically dominant solution. In which case, a complete treatment of the problem would involve perturbing the equations of motion of IIB supergravity directly in ten dimensions and constructing the concomitant black holes directly in ten dimensions. Second, it would be extremely interesting to understand how to reproduce the entropy curves we have constructed using the CFT. One of the consequences of our work is, of course, the fact that the number of parameters match on both sides of the duality. While this is a step towards solving this counting problem, there still remains much work to be done towards resolving the entropy puzzle. Third, $\alpha^\prime$ corrections to the IIB action involving the ten-dimensional metric and five form flux are known \cite{Paulos:2008tn}, and could in principle be used to compute the behaviour of our entropy curves beyond the large t'Hooft limit. Finally, we have not managed to construct the limiting hairy supersymmetric solutions, and instead we approach these solutions from finite temperature. It would be very interesting to find a numerical or analytical procedure that would be able to capture these solutions directly. We are currently investigating whether these elusive black holes can be found by using supersymmetry. It is possible that the diverging tidal forces indicate some non-analytic behaviour at extremality, which would complicate the near horizon expansion\footnote{Note that this is not necessarily the case: Lifshitz spacetimes with critical exponent $z=2$ are analytic in the approach to the horizon, and yet have diverging Tipler integrals.}.

\section*{Acknowledgments}

It is a pleasure to thank N.~Dorey, G.~T.~Horowitz, J.~Maldacena, S.~Minwalla and H.~S.~Reall for discussions. We would like to thank C.~V.~R. Board and \'O.~J.~C.~Dias for reading an earlier version of this manuscript. JES was supported in part by STFC grants PHY-1504541 and ST/P000681/1. JM was partly supported by an STFC studentship.
 
\appendix
\section{\label{sec:appendixA}Equations of motion and conserved quantities}

In this section, we present the equations of motion derived from the action studied in the main paper. The Einstein, Maxwell and scalar equations are as follows
\begin{subequations}
\label{eq:a}
\begin{align}
\label{eq:a:eeq}
&R_{ab}-\frac{1}{2}g_{ab}R-6 g_{ab}=\frac{3}{2}T_{ab}^{EM}+\frac{3}{8}T_{ab}^{mat}\,,\\
\label{eq:a:maxwell}
&\nabla_b F^{b}_{\phantom{a}a}-\frac{1}{4}\varepsilon_{acdef}F^{cd}F^{ef}=\frac{i}{4}\left[\Phi^* (D_a\Phi)-\Phi(D_a\Phi)^*\right]\,,\\
\label{eq:a:scalar}
&D_a D^a\Phi+\Phi\left[\frac{(\nabla_a\lambda)(\nabla^a\lambda)}{4(4+\lambda)^2}-\frac{\nabla_a \nabla^a \lambda}{2(4+\lambda)}+4\right]=0\,,
\end{align}
\end{subequations} 
where the energy-momentum tensor is given by
\begin{align}
\begin{split}
T_{ab}^{EM}&=F_a{}^c F_{bc}-\frac{1}{4}g_{ab}\,F^2,\\
T_{ab}^{mat}&=\frac{1}{2}\left[D_a\Phi\,(D_b\Phi)^*+D_b\Phi\,(D_a\Phi)^*\right]-\frac{1}{2}g_{ab}(D_c\Phi)(D^c\Phi)^*+2g_{ab}\, \lambda\\
&\hspace{+8em}-\frac{1}{4(4+\lambda)}\left[(\nabla_a\lambda)(\nabla_b\lambda)-\frac{1}{2}g_{ab}(\nabla_c\lambda)(\nabla^c\lambda)\right].
\end{split}
\end{align}
\noindent Here $\lambda=\Phi\Phi^*$, and $\phantom{}^*$ denotes complex conjugation. In the radial gauge, this yields a system of seven non-linear differential equations, two of which are of first order, and the rest are of second order.

Here we define conserved charges of the system. The energy $M=N^2m$, angular momentum $J=N^2j$, electrostatic charge $Q=N^2q$ and the electrostatic potential at the boundary $\mu$ are read from the large $r$ asymptotics of the metric functions. Here $N$ is the gauge group rank, and we will work with the rescaled charges $m$, $j$, $q$ and the Bekenstein-Hawking entropy $s=S/N^2$. To compute the conserved charges associated with the asymptotic conformal Killing vector fields ($\partial_t$, $\partial_\psi$), we use the Ashtekar, Das and Magnon formalism~\cite{Ashtekar:1999jx}. The mass is computed with respect to the background AdS$_5$. The electric charge is obtained by computing the flux of the electromagnetic field tensor at infinity
\begin{equation}
Q=\frac{1}{16 \pi G_5}\int_{\mathbb{S}^3_{\infty}}(\star F -F\wedge A)\,,
\end{equation}
and since the magnetic field asymptotically vanishes, the Chern-Simons term doesn't contribute. These quantities have to satisfy the first law of thermodynamics
\begin{equation}
\mathrm{d}m=T\mathrm{d}s+3\mu\mathrm{d}q+2\Omega\mathrm{d}j\,,
\label{eq:firstlaw}
\end{equation}
\noindent where in the non-rotating frame at infinity $\mu$ is the conjugate potential for the electric $U(1)$ charge, $\Omega$ is the thermodynamic rotational potential~\cite{Gibbons:2004ai}, and $T$ is the Hawking temperature. At the highest resolutions, our solutions satisfy the first law to better than $0.001\%$ accuracy.

\section{\label{sec:appendixB}Tidal forces}

Static charged, near extremal black holes can exhibit diverging tidal forces as measured by a freely infalling observer, while having all curvature scalars finite at the horizon~\cite{Horowitz:1997uc}. In order to test whether the near extremal hairy solutions possess the tidal force singularity, we need to analyse the geodesic motion in these backgrounds. We start by considering the metric \textit{ansatz} as given in the main paper, and look for radial, timelike ingoing geodesics parametrised by the proper time $\tau$ and with the tangent vector $\dot{X}^a=\mathrm{d}X^a/\mathrm{d}\lambda$. The Killing vector fields $\partial_t$, $\partial_\psi$ and $\partial_\phi$ give us three conserved quantities
\begin{align}
E=-g_{t a}\dot{X}^a,\qquad L_\psi=g_{\psi a}\dot{X}^a,\qquad L_\phi=g_{\phi a}\dot{X}^a,
\end{align}
\noindent and we consider radial static geodesics on the $S^2$ with $\dot{\theta}=0$ and $\dot{\phi}=0$, and zero angular momentum in the $\psi$ direction. Using the normalization condition ${\dot{X}^a\dot{X}_a=-1}$ we obtain

\begin{equation}
\dot{X}^a=\left\{\frac{E}{f(r)},-\frac{\sqrt{E^2-f(r)}}{\sqrt{f(r)g(r)}}, E\frac{\Omega(r)}{f(r)},0,0\right\},
\label{eq:geodesic}
\end{equation}
\noindent where the coordinates are ordered as $\{t, r, \psi,x, \phi\}$. Here we use the angular coordinate $x=\cos{\theta}$ for numerical convenience.

In order to compute the curvature measured by a freely falling observer along the radial timelike ingoing geodesic, we change into a parallely propagated orthonormal frame (PPON). In the PPON, we require $(\tilde{e}_0)_a=\dot{X}_a$. We choose
\begin{equation}
\begin{split}
(\tilde{e}_0)_a&=-E\,\partial_a t-\sqrt{E^2\frac{g(r)}{f(r)}-g(r)}\,\partial_a r\,,\\
(\tilde{e}_1)_a&=\sqrt{E^2-f(r)}\,\partial_a t-E\sqrt{\frac{g(r)}{f(r)}}\,\partial_a r\,,\\
(\tilde{e}_2)_a&=\frac{1}{2}\sqrt{1-x^2}r\,\partial_a \phi\,,\\
(\tilde{e}_3)_a&=\frac{1}{2}\sqrt{\frac{1}{1-x^2}}r\,\partial_a x\,,\\
(\tilde{e}_4)_a&=-\sqrt{h(r)}r\Omega(r)\,\partial_a t+\sqrt{h(r)}r\,\partial_a \psi+\frac{1}{2}x\sqrt{h(r)}r\,\partial_a \phi\,,
\end{split}
\end{equation}
\noindent which satisfies the orthonormality condition
\begin{equation}
g^{ab}(\tilde{e}_\alpha)_a(\tilde{e}_\beta)_b=\eta_{\alpha\beta}\,.
\end{equation}
 The components of the Riemann tensor in the PPON frame are related to the components in the coordinate frame by
\begin{equation}
R_{\alpha\beta\gamma\delta}=R_{abcd}(\tilde{e}_\alpha)^a(\tilde{e}_\beta)^b(\tilde{e}_\gamma)^c(\tilde{e}_\gamma)^d\,,
\end{equation}
\noindent and are diverging, therefore exhibiting a parallely propagated curvature singularity.

\section{\label{sec:appendixC}Numerical convergence}

\begin{figure}[!t]
\centering
    \begin{minipage}[t]{0.45\textwidth}
    \includegraphics[width=\textwidth]{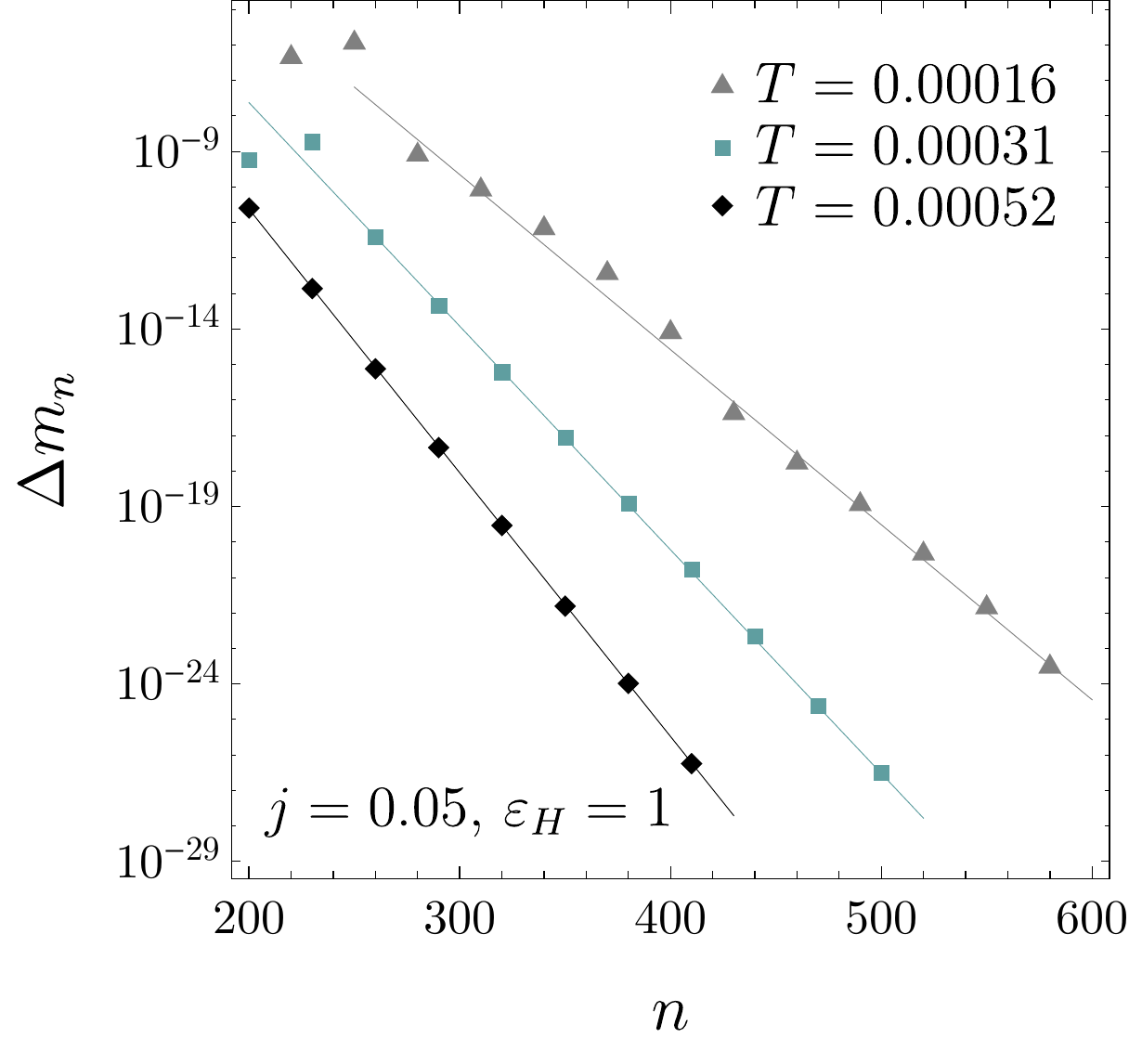}
  \end{minipage}
  \hfill
      \begin{minipage}[t]{0.45\textwidth}
    \includegraphics[width=\textwidth]{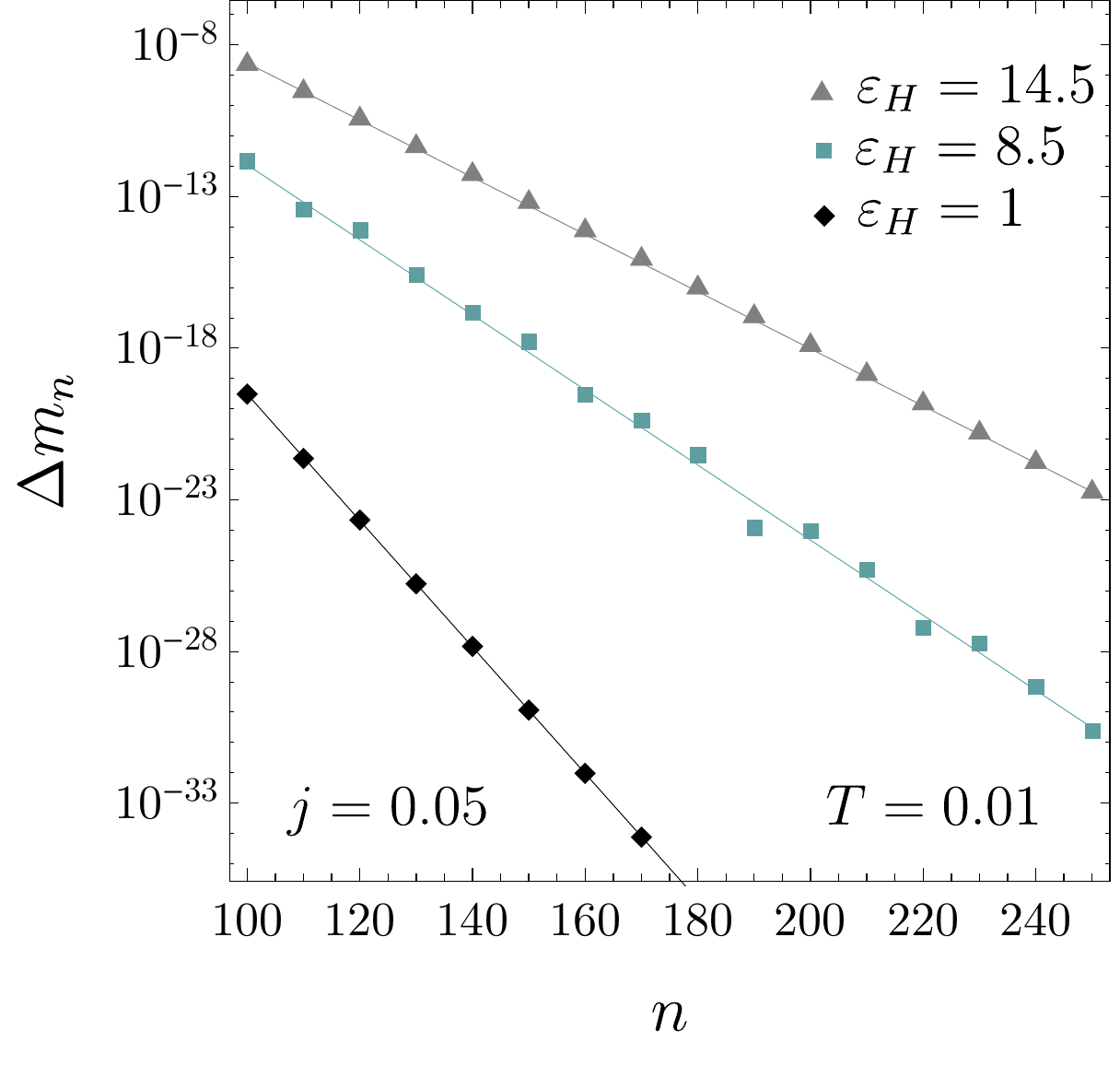}
  \end{minipage}
  \caption{\textit{Left}: Convergence of the energy for a few coldest $J=0.05$, $\varepsilon_H=1$ hairy black holes considered in the paper, with $T=0.00052$ (black rhombi), $T=0.00031$ (blue squares) and $T=0.00016$ (gray triangles). Here we plot the fractional error $\Delta$ against the grid size $n$. \textit{Right}: Low temperature ($T= 10^{-2}\pm 10^{-3}$) hairy black hole energy convergence for a few different horizon scalar values $\varepsilon_H$. Black rhombi are for $\varepsilon_H=1$, blue squares for $\varepsilon_H=8.5$, and gray triangles for $\varepsilon_H=14.5$. Here we show the fractional error against the grid size in a log scale.}
 \label{fig:convergence}
\end{figure}

We perform numerical calculations in the radial gauge, which is allowed by the fact that our problem is reduced to be of co-homogeneity one. This turns out to be significantly advantageous over the commonly employed DeTurck gauge, which we find that is not well behaved in the near extremal limit. We discretize the equations using a pseudospectral collocation method on a Chebyshev grid, and solve the resulting numerical equations using the Newton-Raphson algorithm.

We find that in the radial gauge the near-extremal rotating hairy black holes exhibit exponential convergence. In Fig.~\ref{fig:convergence}, we present convergence test for the black hole energy, plotting the fractional error
\raggedbottom
\begin{equation}
\Delta m_n=\left| 1- m_{n+1}/m_n\right|\,
\end{equation}
\noindent against the grid size $n$. The convergence worsens significantly at very low temperatures (Fig.~\ref{fig:convergence} \textit{left}), large horizon scalar fields $\varepsilon_H$ (Fig.~\ref{fig:convergence} \textit{right}) and large angular momenta, due to large gradients in the functions.
\\\\

\section*{References}
\bibliographystyle{iopart-num}
\bibliography{refs}{}
\end{document}